\documentclass{aa}
\usepackage{graphics}
\usepackage{graphicx}
\usepackage{amssymb}
\usepackage{astron}
\begin{document} 
\newcommand{\ansteebarklem}{Anstee \& O'Mara (1995), 
Barklem \& O'Mara (1997) and Barklem et al. (1998)}
\newcommand{\barklem}{Barklem \& O'Mara (1997)}
\newcommand{\barklemetal}{Barklem et al. (1998)}
\newcommand{\carlsson}{Carlsson (1986)}
\newcommand{\drakenlte}{Drake (1991)}
\newcommand{\edvardsson}{Edvardsson et al. (1993)}
\newcommand{\feltzinggustafsson}{Feltzing \& Gustafsson (1998)}
\newcommand{\kurucz}{Kurucz et al. (1984)}
\newcommand{\moore}{Moore (1966)}
\newcommand{\ols}{Olsen (1984)}
\newcommand{\olsi}{Olsen (1993)}
\newcommand{\olsii}{Olsen (1994)}
\newcommand{\olsiii}{Olsen (1995)}
\newcommand{\smith}{Smith (1981)}
\newcommand{\thorenletter}{Thor\'en (2000)}
\newcommand{\thorennlte}{Thor\'en (2000, in preparation)}
\newcommand{\thorenfeltzing}{Thor\'en \& Feltzing (2000)}
 
\title{Chemical abundances in cool metal rich disk dwarf stars
\thanks{Based on observations made at ESO, La Silla}} 
\titlerunning{Abundances in cool metal rich dwarfs}
\thesaurus{7(08.01.1,08.12.1)}
\offprints{Patrik Thor\'en}
\author{Patrik Thor\'en \inst{1} and Sofia Feltzing \inst{2}}
\date{Recieved 5 June 2000 / Accepted 17 September 2000}
\institute{Uppsala Astronomical Observatory, Box 515, S-751 20, Uppsala, Sweden
\and Lund Observatory, Box 43, S-221 00 Lund, Sweden}
\maketitle
%
%
\begin{abstract}

The present study of spectra of twelve metal-rich cool dwarf stars,
carefully selected in order to cover a range of temperatures 
($\sim 4400 - 6000$ K), is a
follow up on Feltzing \& Gustafsson (1998)\nocite{feltzing} with the aim to understand
the apparent over-ionization and anomalous elemental abundances found
by them for the  K dwarf stars in their sample.

Our method of analysis employs synthetic spectra of the full spectrum
both to constrain the continuum level and to derive abundances. It is shown that
by using this method and  imposing a strict excitation equilibrium
(possible to do because of the  care in selection of observed Fe {\sc i}
lines) we are able to show that metal-rich K dwarf  stars do not
show anomalous stellar abundances, as indicated in Feltzing \&
Gustafsson (1998), and can, with reasonable efforts, be analyzed in order to
increase the number of metal-rich stars with useful chemical
abundances.

With abundance analysis  by means of spectrum synthesis and
assuming Local Thermodynamic Equilibrium (LTE) the abundances of Na,
Si, Ca, Sc, Ti, V, Cr, Fe, Co, Ni, and Nd have been derived.  Also
ionization balance is satisfied for Fe and Cr after correcting the
stellar effective temperatures such that both ionization and excitation
equilibrium were satisfied.

In addition, spectra from five cool dwarf stars of the Feltzing \& 
Gustafsson (1998) sample have been
analyzed with the methods used in this work. They show essentially the
same  abundance patterns as the new stars in this sample.
 

\end{abstract}


%
%
\section{Introduction}

The atmospheres of dwarf stars retain, except for a small number
of elements, the chemical composition of
the gas they were formed out of. This makes them ideal as tracers of
the galactic chemical evolution. Combining chemical and kinematical
data  is a powerful method in studies of the galactic chemical
evolution.  The last decade have seen several studies of stellar
abundances of large numbers of dwarf stars with known kinematics and
derived ages, eg.  Edvardsson et al. (1993) (hereafter
EAGLNT93)\nocite{eaglnt:93}, Feltzing \& Gustafsson (1998) (FG98), 
Fuhrmann (1998)\nocite{fuhrmann} and Chen et
al. (2000)\nocite{chen00}.  These studies have mainly
been concerned with the warmer F and G dwarf stars and the general
trends of the different elements are now well established and their
gross features can be  reproduced, for most elements, by models of
galactic chemical evolution, eg.  Samland
(1998)\nocite{samland} and Timmes et al. (1995)\nocite{timmesetal}. 
Well defined abundance trends for
stars more metal rich than the Sun can also provide vital clues to the 
production sites of several elements (see Samland, 1998).

The evolution above ${\rm [Fe/H]}\sim 0.1$ dex is still, however, fairly
poorly sampled and several elements show large scatter. 
One reason for this is the relative 
rarity of the most metal-rich stars. FG98 studied
a sample of 47 F, G and K dwarf stars with metallicities above solar. In
general they confirmed the trends found in EAGLNT93,
now also confirmed in Chen et al. (2000). However, for the cool K dwarf stars
in their sample FG98 found both an apparent 
over-ionization as well as odd abundances for several elements.
\thorenletter\nocite{thoren00calletter} \ has shown that the odd Ca 
abundances were not, as originally
thought, due to NLTE effects (see also Drake, 1991), but wrongly calculated
damping parameters. The other effects, though, remain to be explained. 

Sect. \ref{sect:obs} describes the observations. Sect. \ref{sect:red}
reports on the data reduction and analysis. In sect. \ref{sect:abund} the
stellar abundances are presented. This is followed by a
discussion in sect. \ref{sect:disc} and conclusion in sect. \ref{sect:conclusion}.

%
%
\section{Observations}
\label{sect:obs}

\subsection{Selection of objects}

In order to, as efficiently as possible, study the behaviour of Ca and 
departures from ionizational equilibrium in metal-rich dwarf
stars we have selected stars with a wide
range of effective temperatures ($\sim 4400 - 6000$ K).

In our selection of stars we used stellar parameters derived from Str\"omgren 
photometry using the calibration in \ols. 
The stars are not distant and interstellar
reddening is therefore expected to be negligible and was not taken
into account.  The data for the ten stars selected in this way are presented
in the first part of Table \ref{tab:starspectable}. Three of these
stars, HD 32147, HD 61606A and HD 69830, were previously studied by FG98.
Two other dwarf stars,  HD 4307 (HR 203) and HD 20807
(HR 1010), studied in EAGLNT93, 
 were chosen to provide a consistency check on both reductions as well
as the subsequent abundance analysis.  The data for these two stars
are presented in the second part of Table \ref{tab:starspectable}.

HD 26441 was originally selected as a program star, however it soon became
clear that this is a doubled lined binary star. This was not known at the
time of selection, but has later been confirmed
by  Martin \& Mignard (1998)\nocite{martinmignard}. 
Attempts have been made to determine 
the abundances for this double star, but the uncertainties proved
to be too large. However, the spectra show that the masses must
be rather similar. Assuming a similar metallicity  ([Fe/H]=0.15) of the two components
the difference in masses (which give the temperature and radius) suggest
one component of solar mass and one of slightly lower mass. 
This is consistent with the results by  Martin \& Mignard (1998), who find
$M_1 =1.036 \pm 0.209~M_\odot$ and $M_2=0.824\pm0.168~M_\odot$.

Finally, we also include a re-analysis of stellar spectra for five dwarf stars,
 originally taken for FG98. These are presented in the 
third part of Table \ref{tab:starspectable}.

\subsection{Selection of wavelength regions}

The Long Camera on the CES only allows observation of one short wavelength 
interval at a time, typically a few times 10 {\AA}. This means that we had to make a careful
selection of wavelength regions. 
Seven wavelength regions were selected such that a maximum number of unblended
strong and weak Ca lines were observed.

\begin{table*}
\caption{\label{tab:starspectable}  Stellar
 model parameters, both as first derived
from photometry and as derived from the stellar spectra themselves
after iteration  (see Sect. \ref{sect:par_det_desc}) 
and used in the final  abundance analysis. 
 Resulting 
 abundances are presented in Table \ref{tab:ab_all}. 
G=typical Gaussian profile FWHM
convolved with models to fit observed spectra. 
 $\log g_{\pi}$ = surface gravities calculated from Hipparcos parallaxes,
see sect. \ref{sect:par_det_desc}.
The two stars in section 2 of the table are  the hot metal poor dwarfs in 
common with EAGLNT93.  The parameters showed are the same as EAGLNT93 used 
in their analysis.
The last five stars are parameters of FG98 stars -  the 2.7m McDonald 
spectra from that sample  have been analysed with the methods described in this work.}
\begin{tabular}{l l r r r r r r l l l }
\hline\noalign{\smallskip}
 && \multicolumn{3}{l}{Photometry} & \multicolumn{3}{l}{Spectroscopy}\\
ID       &  & $T_{\rm eff}$	& $\log g$	& [Me/H] & $T_{\rm eff}$ & $\log g$ & [Me/H] & $\xi$  & G & $\log g_{\pi}$ \\
\noalign{\smallskip}
\hline\noalign{\smallskip}
HD 12235 & & 5971 & 4.18 & 0.15 & 5971 & 4.18 & 0.15 & 1.1 & 7.5  & 4.07 \\
HD 21197 & & 4457 & 4.59 & 0.13 & 4657 & 4.59 & 0.27 & 0.6 & 5.0  & 4.55 \\
HD 23261 & & 5132 & 4.44 & 0.06 & 5132 & 4.64 & 0.06 & 0.9 & 4.5  &  \\
HD 30501 & & 5174 & 4.54 & 0.13 & 5174 & 4.54 & 0.13 & 1.0 & 5.0  &  4.47 \\
HD 31392 & & 5390 & 4.29 & 0.06 & 5390 & 4.59 & 0.06 & 0.7 & 5.0  & 4.40 \\
HD 32147 & & 4625 & 4.57 & 0.17 & 4825 & 4.57 & 0.29 & 0.7 & 4.5  & 4.48 \\
HD 42182 & & 4917 & 4.54 & 0.25 & 5117 & 4.54 & 0.10 & 0.9 & 5.0  & 4.46 \\
HD 61606A & & 4833 & 4.55 & 0.06 & 4833 & 4.55 & 0.06 & 1.0 & 5.0 & 4.46 \\
HD 69830 & & 5484 & 4.34 & 0.10 & 5484 & 4.34 & 0.10 & 1.2 & 5.3  & 4.45 \\
HD 213042& & 4560 & 4.58 & 0.06 & 4760 & 4.58 & 0.19 & 1.0 & 4.5  & 4.51 \\
\noalign{\smallskip}
\hline\noalign{\smallskip}
HD 4307  &HR 203 & & & & 5809 & 4.06 & -0.38 & 1.7 & 5.5  & 3.88  \\
HD 20807 &HR 1010& & & & 5889 & 4.41 & -0.26 & 1.3 & 5.5  & 4.36 \\
\noalign{\smallskip} 
\hline\noalign{\smallskip}
HD 77338  & & 5290  & 4.50  & 0.45  &  5290 & 4.60  & 0.22  & 1.0  & 4.5  & 4.37    \\  
HD 87007  & & 5300  & 4.50  & 0.43  &  5300 & 4.40  & 0.27  & 1.0  & 4.5  & 4.41    \\  
HD 103932 & & 4510  & 4.58  & 0.21  &  4510 & 4.28  & 0.16  & 1.0  & 4.0  & 4.49    \\  
HD 131977 & & 4585  & 4.58  & 0.18  &  4585 & 4.58  & 0.04  & 1.0  & 3.5  & 4.50    \\  
HD 136834 & & 4765  & 4.56  & 0.23  &  4765 & 4.17  & 0.19  & 1.0  & 5.0  & 4.39   \\  
 \noalign{\smallskip}
\hline
\end{tabular}
\normalsize
\end{table*}

\begin{table*}
\caption{\label{tab:observedregions}  Observed wavelength regions. 
Asterisks indicate that a star was observed in that wavelength region. 
 The stars are arranged in the same way as in Table \ref{tab:starspectable}}.
\begin{tabular}{l l l l l l l l}
  \noalign{\smallskip}
\hline\noalign{\smallskip}
Star     & 5248  & 5800  & 6105  & 6162  & 6448   & 6798   & 8520       \\
HD 12235 &       & \ \ * &       & \ \ * & \ \ *  & \ \ *  & \ \ *      \\
HD 21197 & \ \ * & \ \ * & \ \ * & \ \ * & \ \ *  & \ \ *  & \ \ *      \\
HD 23261 & \ \ * & \ \ * & \ \ * & \ \ * & \ \ *  & \ \ *  & \ \ *      \\
HD 30501 & \ \ * & \ \ * & \ \ * & \ \ * & \ \ *  & \ \ *  & \ \ *      \\
HD 31392 & \ \ * & \ \ * & \ \ * & \ \ * & \ \ *  & \ \ *  & \ \ *      \\
HD 32147 & \ \ * & \ \ * & \ \ * & \ \ * & \ \ *  & \ \ *  & \ \ *      \\
HD 42182 &       & \ \ * &       & \ \ * & \ \ *  & \ \ *  & \ \ *      \\
HD 61606A & \ \ * & \ \ * & \ \ * & \ \ * & \ \ *  & \ \ *  & \ \ *      \\
HD 69830 & \ \ * & \ \ * & \ \ * & \ \ * & \ \ *  & \ \ *  & \ \ *      \\
HD 213042&       & \ \ * &       & \ \ * & \ \ *  & \ \ *  & \ \ *      \\
The Sun  & \ \ * & \ \ * & \ \ * & \ \ * & \ \ *  &        & \ \ *      \\ 
\noalign{\smallskip}
\hline\noalign{\smallskip}
HR 203   & \ \ * & \ \ * & \ \ * & \ \ * & \ \ *  & \ \ *  & \ \ *      \\       
HR 1010  & \ \ * & \ \ * & \ \ * & \ \ * & \ \ *  & \ \ *  & \ \ *      \\       
\noalign{\smallskip}
\hline\noalign{\smallskip}
HD 77338 & \ \ * & \ \ * & \ \ * &       & \ \ *  & \ \ *  &            \\ 
HD 87007 & \ \ * & \ \ * & \ \ * &       & \ \ *  & \ \ *  &            \\ 
HD 103932& \ \ * & \ \ * & \ \ * &       & \ \ *  & \ \ *  &            \\ 
HD 131977& \ \ * & \ \ * & \ \ * &       & \ \ *  & \ \ *  &            \\  
HD 136834& \ \ * & \ \ * & \ \ * &       & \ \ *  & \ \ *  &            \\           
\hline\noalign{\smallskip}
\end{tabular}
\normalsize
\end{table*}

The choice of wavelength regions was made 
so as to maximize the numbers of useful (for abundance determination)
 lines from elements  other than  Ca. In particular
we wanted to cover a large range in excitation potential for Fe {\sc i} lines
(this later proved vital for our analysis)  as well as  Fe {\sc ii}, Ni {\sc i}, Cr {\sc i} and Cr {\sc ii}. 

We were furthermore guided  in our selection by the solar line table by \moore\nocite{mooretab} and
the solar atlas by \kurucz\nocite{solaratlas}. Our final seven wavelength regions 
are centered at  5248, 5880, 6105, 6162, 6445, 6798, and 8520 {\AA}. 
 For spectral coverage of single stars, see Table \ref{tab:observedregions}.
Note, for the five re-analysed stars, that these spectral regions were
not analysed in FG98.

\subsection{Instrumental setup and observations}
\label{sect:instrument}
The spectra were obtained with the ESO CAT-CES November 19 - December 1 
1995. The Long Camera and {\sc ccd} 38 were used, with a nominal
spectral resolution of 70,000-100,000, depending on wavelength
region. The true resolution is slightly lower in the blue due to  
bleeding in the {\sc ccd}.  Typical S/N ratios were between 150 and 200.

Spectra from five stars of the sample of FG98 were also analysed 
with the methods described in this work. The stars were
observed with the 2.7 m telescope at the McDonald Observatory, University
of Texas, spring 1994. The results of this subsample should be looked 
upon with caution since we have not managed
to analyse all the lines used in the ESO CAT sample and also since it has been 
observed with a different instrument.  The description of the McDonald 
observation run can be found in FG98.

%
%
\section{Reductions and abundance analysis} 
\label{sect:red}

\subsection{Basic reduction steps}

The spectra were reduced with the 
ESO MIDAS \cite{midas} software. First
the bias was subtracted from all science and calibration frames. Then the 
flatfield images were coadded and all science frames divided by these, 
normalized, flatfields. The stellar and calibration spectra were then
extracted from the frames. 

Calibration spectra, from a Th-Ar lamp, were obtained for each wavelength
setting.   The lines in the Th-Ar spectrum were identified using a 
 Th-Ar atlas \cite{tharg_atlas}. Finally the stellar spectra were 
wavelength calibrated using the LONG context in MIDAS.

A spectrum of a cool dwarf star is rich in lines which makes the definition
of the continuum difficult in many cases. Therefore we did not only
use continuum points identified in a solar atlas \cite{solaratlas}
and the Arcturus Atlas by Griffin (1968)\nocite{griffin_arcturus} \  but also
calculated synthetic stellar spectra to define the exact continuum. 

The final continuum level was decided upon through an iterative process were
also the stellar parameters were adjusted to obtain optimal fits to
the continuum as well
as retaining  excitational equilibrium for Fe {\sc i}.

\begin{figure*}                                           
  \resizebox{160mm}{!}{\includegraphics{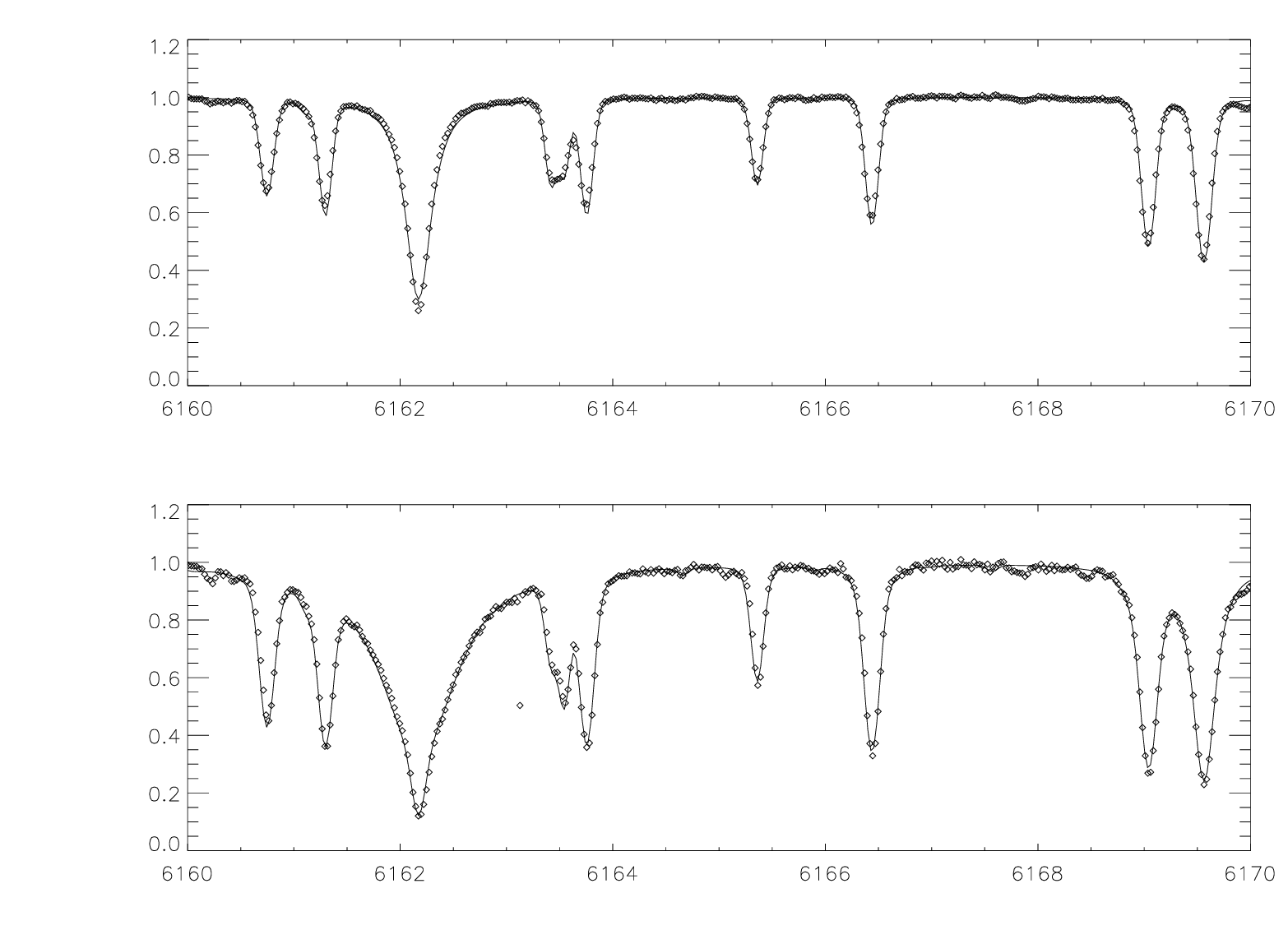}}
  \hfill
  \parbox[b]{160mm}{
    \caption{CAT spectrum (diamonds) and synthetic spectrum (solid)
     of the Sun (upper) and HD 61606A (lower).}
    \label{fig:synth_obs_example}}
  \end{figure*}

An example of the quality of our spectra as well as the goodness of
the fit of the synthetic spectra to the observed ones is shown 
in Fig. \ref{fig:synth_obs_example}.

\subsection{Scattered light}

Due to scattered light from an unknown source in the {\sc ccd} frames, 
a background varying in both dispersive and cross-dispersive direction
 had to be removed from them.
 This background had a similar pattern as 
the fringing pattern which is normally seen in the red, 
but with a lower amplitude and sometimes slightly shifted  in one or
both directions. It was also
time-dependent and therefore its effect on the spectra for the fainter stars
was very strong. A polynomial function was fitted to the smoothed 
background with the MIDAS SKYFIT/LONG procedure. 

After background subtraction  the reduced spectra became considerably cleaner. 
Without removal of the background the patterns   
were still visible after division with flatfields, since the flatfields
were  taken with a short exposure time, not long enough to produce the 
scattered light background component.

%
%
\label{sect:lte}

\subsection{Stellar parameters}

Initial stellar parameters were determined from Str\"omgren photometry.
The colours were taken from the catalogues in \olsi\nocite{olsen1993} 
and \olsii\nocite{olsen1994}. We used the calibration in \ols,
 which is valid for G to K dwarf stars.

\begin{eqnarray}
\log T_{\rm eff} & = & 0.341(b-y) + 3.869 \nonumber\\
\log g     & = & 0.37(b-y) + 4.35  \nonumber\\
{\rm [Me/H]}        & = & -5.1 dm_1 + 0.07 
\label{stromgrenrelation1}
\end{eqnarray}

\noindent
for stars with $b-y>0.51$ and

\begin{eqnarray}
\log T_{\rm eff}  & = & 0.469(b-y) -0.103 dm_1 + 0.087 dc_1 + 3.947 \nonumber\\
\log g      	  & = & 0.50(b-y) + 4.10 dm_1 - 2.56 dc_1 +4.30 \nonumber\\
{\rm [Me/H]}      & = & -7.5 dm_1+1.9 dc_1-0.04
\label{stromgrenrelation2}
\end{eqnarray}
 
\noindent
for stars with $b-y<0.51$.

The stellar parameters were subsequently changed for some of the stars,
see Sect. \ref{sect:par_det_desc} and Table \ref{tab:starspectable}.

\subsection{Model atmospheres} 
We used the MARCS code,  first described by Gustafsson et
al. (1975)\nocite{gustafsson_marcs}, to generate the model atmospheres.  
Since then the
program has been further developed in various ways and updated in
order to handle the line blanketing of millions of absorption lines
more accurately, see EAGLNT93 and Asplund et al. (1997)\nocite{asplund_marcs}.
In particular, these models reproduce the 
continuous opacity and ionizing radiation field in UV better.

\subsection{Atomic line data}
Basic line data were taken from the VALD database 
\cite{vald}.  For Ca we used oscillator strengths from \smith \ 
and, for Ca 6162, from Smith \& O'Neill (1975). \nocite{smithoneill:75}

For other elements we derived log $gf$ values both for the lines in Table
\ref{tab:linetable} and lines (not used in the abundance analysis) in the
wings of strong lines.
Astrophysical oscillator strengths were derived from
 the observed solar spectra. For the 6798 \AA \ region 
no spectrum was achieved for the Sun. A Kurucz solar spectrum 
was used for that region. 
The initial values from VALD were usually within the same order of magnitude
as the astrophysically fitted values.

Atomic line pressure damping parameters from \ansteebarklem
 \  were calculated with  the Uppsala code SPECTRUM and 
used for all elements except Ca. These damping parameters 
 give a good fit. The damping calculations are not available for all 
transitions. For those transitions a $\delta\Gamma_6$ damping value typical 
for that element was inserted. These are shown in Table \ref{tab:linetable}. 
One of the aims of the project was the comparison of
LTE and NLTE analysis for Ca (Thor\'en, 2000, in preparation). 
The pressure damping treatment 
used in SPECTRUM (Anstee \& Barklem, 1997 and Barklem, 1998) is not available
for MULTI \cite{multi} (the code used for the NLTE calculations, see Thor\'en, 2000, in preparation), 
therefore  pressure damping data for Ca were 
taken from \smith \ and O'Neill \& Smith (1980) \nocite{oneillsmith:80} to achieve maximum consistency with 
the NLTE analysis. 
 
\subsection{Abundance determinations} 
\label{sect:par_det_desc}

To derive stellar abundances we used standard LTE calculations. The
procedure differs in some important ways from that of EAGLNT93
and FG98 and therefore merits a separate discussion.
We now briefly describe our iterative process for finding stellar parameters
and determine the exact continua in difficult regions:

The instrumental profile was determined by 
adjusting the width of
a Gaussian convolved with the synthetic spectrum  generated with SPECTRUM.
 In the NLTE study 
(Thor\'en, 2000, in preparation) the same Gaussian width is used. 
For some stars the Gaussian width
 varied with a small amount ($< 10 \% $) from region to region. 
The Gaussian profile simultaneously accounts for the instrumental profile
and any effect of stellar rotation and macroturbulence.
The Gaussian FWHM width varied between  4.5 and 8.5 km/s (except the spectra
from the sample of FG98, some of which had FWHM down to 3.5 km/s). 
Typical Gaussian widths for each star is given in 
Table \ref{tab:starspectable}.  

Observed stellar spectra were compared with synthetic spectra generated with 
SPECTRUM. To estimate corrections to the stellar parameters derived from the 
Str\"omgren photometry, equivalent widths
were obtained from the fitted synthetic spectra. This 
was done for the three wavelength regions centred at 6162, 6455, and
6798 {\AA}. These regions are the same as those containing
Ca lines in FG98. Using derived abundances for Fe lines and their
 equivalent widths measured in the synthetic spectra we could
constrain effective temperature and microturbulence
by imposing excitation balance (effective temperature)
and minimizing the spread in derived abundance (microturbulence). 

\begin{figure}                                           
  \resizebox{80mm}{!}{\includegraphics{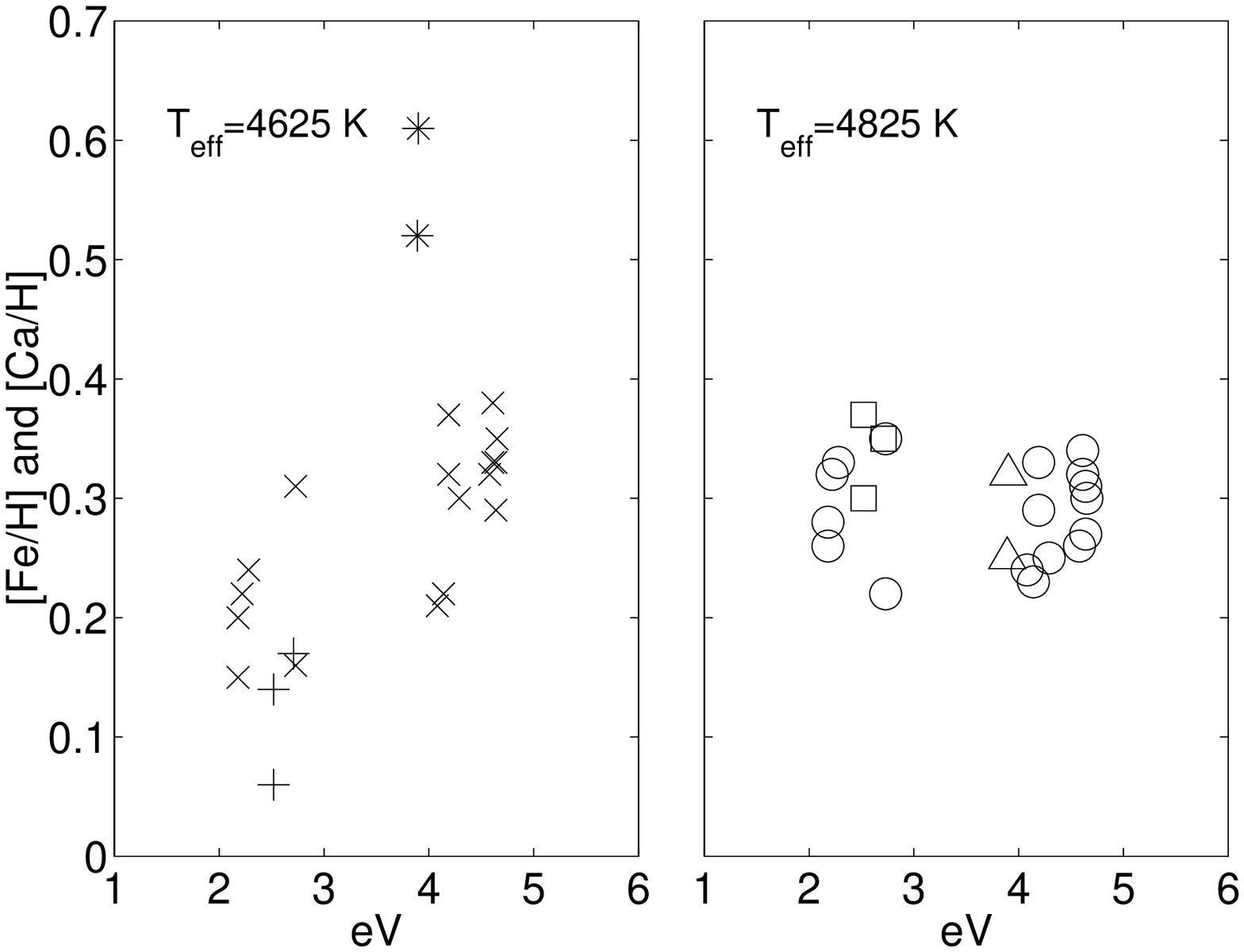}}
  \hfill
  \parbox[b]{80mm}{
    \caption{Abundances for different Fe {\sc i}, Fe {\sc ii} and and Ca lines of HD 32147 versus the 
    excitation energy of the lower level. $\times$, $ * $ and $ + $ : Fe {\sc i}, Fe {\sc ii}  and Ca abundances 
  before temperature correction. $ \bigcirc $, $ \triangle $ and $ \square $ : Fe {\sc i} , Fe {\sc ii} 
  and Ca abundances after temperature correction. Note the great Fe {\sc ii} 
  overabundance for the unmodified temperature model.}
    \label{fig:exc_balance}}
  \end{figure}

An example, for HD 32147, of corrections with respect to excitation balance is 
given in  Fig. \ref{fig:exc_balance}. The effective temperature 
 was raised from 4625 K to 4825 K in order to reach excitation balance. 
In total four of our stars needed temperature adjustments. They were among the
coolest stars in the sample and the adopted correction  was in all cases +200 K,
see Table \ref{tab:starspectable}.

In the next step, the other spectral regions were analyzed 
with synthetic spectra.

\begin{figure*}                                           
  \resizebox{160mm}{!}{\includegraphics{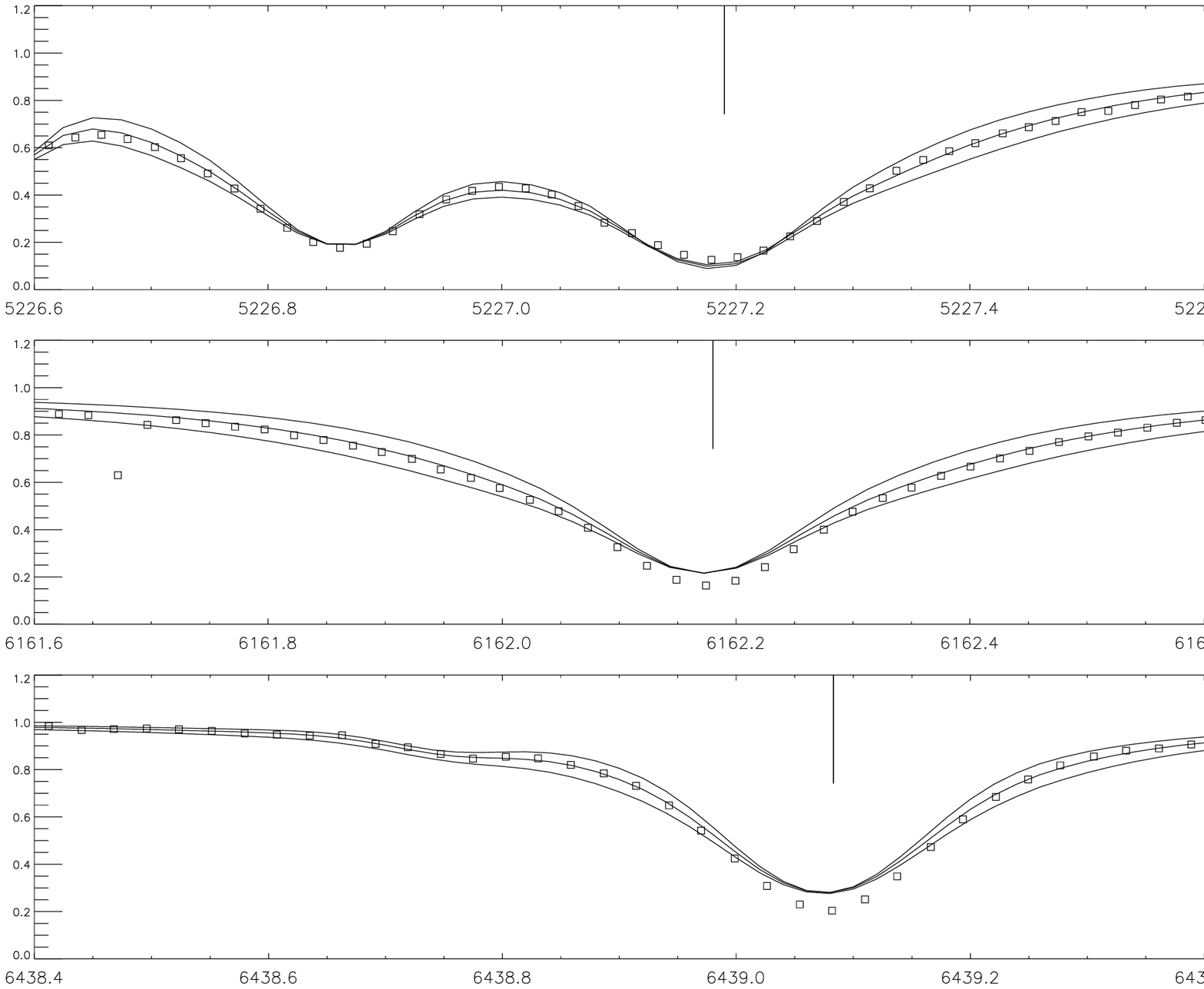}}
  \hfill
  \parbox[b]{160mm}{
    \caption{Example of how the surface gravity was fitted, by 
    using several strong lines. Unit on x-axis is \AA ngstr\"om. 
    The normalized flux data (squares) is from the star 
    HD 31392. Three different model gravity spectra are
    plotted : log g = 4.29, 4.59 and 4.89. The strong lines displayed are
    Fe I 5227.19 \AA, 
    Ca  I 6162.17 \AA \ and Ca  I 6139.07 \AA.
    The final surface gravity chosen was log g=4.59.}
    \label{fig:loggfit}}
  \end{figure*}

The surface gravity was first determined
by fitting the pressure broadened Ca {\sc i} line at 6162 \AA.
However, when this line was well fitted the quality of the fit of
the synthetic spectra in the other regions was not always optimal. 
For the cool K dwarf stars there are many more lines that are affected
by changes in $\log g$. We therefore decided to fit  the surface gravity by 
optimizing the fits of strong lines in all wavelength regions. Examples of the
fitting are shown in Fig. \ref{fig:loggfit}.

The effective temperature
 and surface gravity parameters were then changed in order to achieve
 excitation balance in Fe (effective temperature,  now using the
 full spectral coverage) and good fits,
  in the synthetic spectra, of the wings of strong  lines (surface gravity). 
The choice of surface gravity from synthetic spectral analysis 
seems to be appropriate also when examining the ion/neutral ratios in
Fig. \ref{fig:ion_eq} where the signs of overionization is  shown to be moderate.

A comparison with surface gravities derived from parallaxes can now be 
done with the accurate data from Hipparcos \cite{hipparcos}. With the formula 
(e.g. Gustafsson et al. 1974)

\begin{eqnarray}
\log \frac{g}{g_{\odot}}=4 \log \frac{T_{\rm eff}}{T_{{\rm eff} \odot}}+ \log \frac{\cal{M}}{\cal{M}_{\odot}} + 0.4(M_{\rm bol}-M_{\rm bol \odot})
\end{eqnarray}
we get, with $M_{\rm bol \odot}=+4.76$ \cite{mihalasbinney}, 
$\log g_{\odot}=4.44$ and
$T_{\rm eff \odot}=5780$

\begin{eqnarray}
\log g=4 \log T_{\rm eff} + \log \frac{M}{M_{\odot}} +  0.4 V + \nonumber \\
  0.4 B.C. - 2 \log \frac{1}{\pi} - 10.5117
\end{eqnarray}

The parallax surface gravities of the dwarves are presented in 
Table \ref{tab:starspectable}. 
Parallaxes and V-magnitudes were obtained from the Hipparcos catalogue,
which contained data for all objects but HD 23261.
Masses and bolometric corrections were derived from 
Table 3--6 in Mihalas \& Binney (1981) and the 
absolute magnitudes derived from the Hipparcos catalogue. 
Masses for two of the
stars, HD 12235 and HD 69830, have been derived by  
Feltzing et~al. (2000) and these agree within 0.02 dex 
to the masses estimated in this work. 
The parallax gravities agrees in most cases well with the ones derived
by spectroscopy. Note that the largest deviations appears in 
the re-analysed FG98 sample, which contains less spectra for our
abundance analysis.

\subsection{Error budget from the analysis}

\subsubsection{NLTE effects in Fe {\sc i} and Ca {\sc i}}

NLTE abundance errors  for low-excitation Fe lines would be important
especially for the stars that have required temperature changes
to obtain excitation balance in Fe {\sc i} lines; HD 21197, HD 32147,
HD 42182 and HD 213042. These are among the coolest stars, which
may raise some suspicion about NLTE effects for low excited
Fe {\sc i} lines for cool metal rich stars. Holm (1996) and Thevenin \& Idiart
(1999) calculate Fe NLTE correction for stars with parameters close to the cool
dwarf stars in this sample. In both works it seems that NLTE effects 
decrease and disappear when stellar parameters approach the parameters of 
our dwarf stars. Therefore we assume that NLTE effects for Fe {\sc i} can be neglected 
and that our use of Fe {\sc i} line excitation balance may safely be used. 

NLTE effects in Ca have been explored by \thorennlte. They are
shown to be small for cool dwarf stars. For the hotter dwarf stars in this
sample LTE Ca abundance values may be slightly too high. However the effects
increase with line strength and the lines used in this work never get 
strong enough to produce significant errors. 
 
\subsubsection{Line data}

Errors in the atomic line data are restricted to the changes
we have made to fit the ESO CAT solar observations, to the
broadening parameters calculated with the theories presented in 
\ansteebarklem\nocite{anstee,barklem:97,barklem98:1}
  \ and to the laboratory measurements presented by 
\smith. The error in the atomic data from the 
latter source is reported to be small ($<$5 \%). The theoretically
calculated pressure broadening parameters have been tested in 
\ansteebarklem \  by calculating solar abundances. These abundances
are shown to be consistent with meteoritic abundances. In 
the work presented here erroneous damping parameters would be visible when fitting 
synthetic spectra. Therefore we conclude that such errors are 
negligible  for our analysis.

The error in astrophysical 
oscillator strength corresponds to the error in fitting the solar data,
$<$0.05 dex. 
 
\subsubsection{Continuum fitting}

The continuum levels have been drawn with help of synthetic spectra. 
Thus the errors in these are more dependent on the spectral
synthesis than on anything else. Possible errors in the continuum levels
 have been included in the errors from spectral fitting. 
These are deemed to be small ($<$ 0.1 dex) and in cases of
doubts concerning the continuum level, lines have not been included in 
the abundance analysis.

\subsubsection{Background subtraction}

The subtracted scattered light background component could possibly introduce an
error, if the background model is badly represented. This would be
visible as a continuum difficult to define and such errors are included
in the line profile fitting (Sect. \ref{sect:synthfit}).

\subsubsection{Instrumental profile and macroscopic stellar broadening}

The combined line profile broadening by the instrument and any stellar 
macroscopic broadening were represented by a Gaussian profile that was convolved
with the synthetic spectra. This had to be fitted to every new spectral
region and star since the resolving power varies with wavelength.
 The fitting was done by examining
the shapes of all lines in an observed spectrum. 
For most stars and regions the
Gaussian profile had a FWHM of $\sim$ 5.0 km/s. If an erroneous profile 
would be used, the measured abundances would be wrong. The maximum error
from misfitted Gaussian profiles is probably $<$ 0.05 dex, considering
the large number of lines used to determine the instrumental profile.
An  inconsistent FWHM value would be visible in form of a systematic
shift in the abundances for the wavelength region and star concerned.

\subsubsection{Synthetic line profile fitting}
\label{sect:synthfit}

This error source includes uncertainties in the continuum.
In most cases the line fitting uncertainty have been judged to 
0.03-0.1 dex. For strong lines it may be larger. However the abundances
do not change significantly when the stronger lines are excluded.

\subsubsection{Errors in derived abundances due to errors in stellar parameters}

Because of the method employed in the abundance analysis a straightforward 
error analysis done by varying the surface gravity and effective
temperature used in the model atmosphere is not possible. However, 
Feltzing (1995) provided such an analysis for two of the  K dwarf  stars studied
here, HD 32147 and HD 61606A, and we may use their derived changes in stellar
abundances as a function of model parameters as an estimate of the expected
errors in this work.  Furthermore FG98 (their Fig. 7) graphically shows 
how Fe I/Fe II and Cr I/Cr II 
changes with stellar parameter changes for the two K dwarf stars HD 61606A and 
HD 103932.

First they find that the two stars, with quite different metallicities,
all the same respond with roughly the same changes in abundances because 
of the changes in model parameters. 

In particular  Feltzing (1995) find that a change in effective temperature of +200 K 
would decrease Fe {\sc i} with --0.03 dex and Fe {\sc ii} with --0.29 dex
for HD 32147, thus considerably reducing the apparent over\-ionization. Similar effects
were found for HD 61606A and also for Cr. A change in surface gravity of --0.4 dex
would change the Fe {\sc i} abundance with --0.04 dex and the Fe {\sc ii} abundance
with --0.21 dex, also almost erasing the overionization. FG98
found no compelling evidence to change the effective temperatures because they
did not have a large enough span in excitation energy for the Fe {\sc i} lines
to address excitation equilibrium in the coolest K dwarf stars ( they, however,
conclude that "we cannot from our analysis exclude that the apparent 
pattern of overionization ... is due to a temperature scale several hundred
K too low").
Therefore they kept the effective temperatures and a change of 0.4 dex in the surface
gravity  was considered too large. The present study show that indeed for HD 32147 
the effective temperature should be changed by +200 K. 

Further,  Feltzing (1995)  showed that a change of +200 K
meant a change around +0.2 dex for  Na, Ca, Si, Ti, and a -200 K change
-0.2 dex for the same elements.  Si, Cr {\sc ii}, and Fe {\sc ii}
would decrease with 0.15 -0.2 dex for an increase of +200 K, and
increase with the same amount if the change was --200 K.

The effects on Sc {\sc ii}, Co, and Ni are all smaller than 0.1 dex
for changes as large as 200 K.

Since our effective temperatures should have errors smaller or around
100 K we would  expect the errors in derived abundances to be smaller
than the ones quoted above.

\subsubsection{Summary of the error analysis}

The errors that have been suggested to contribute are (with estimated maximum
amplitudes): 
 \begin{tabbing}
\ \ \ \ \ \ \ \ \ \ \ \ \ \ \ \ \ \ \ \ \ \ \ \ \ \ \ \ \ \ \ \ \ \ \ \ \ \=  \ \ \ \=  \\
NLTE effects \> 					: \> $\le 0.05$ dex \\
Line data \> 						: \> $\le 0.07$ dex \\
Instrumental profile and 						 \\
macroscopic broadening \>  	                        : \> $\le 0.05$ dex \\
Synthetic line profile fitting \> 			: \> $\le 0.10$ dex \\
Stellar parameters \> 				        : \> $\le 0.10$ dex 
\end{tabbing}

If the errors are uncorrelated, the worst case would produce an
error of $\sim$ 0.17 dex. However, this is unlikely, considering the
small scatter for most elements seen in Fig. \ref{fig:all_elements}.

\subsection{Comparison with other works}

Two stars, HR 203 and HR 1010 were observed for comparison with the results in
EAGLNT93.  
The differences for elements in common are shown in Table \ref{tab:bdp_diff}.
The differences are mostly small in comparison with the errors
estimated in the two analyses. Notably, there is a difference in Fe (in spite
of the large number of neutral Fe lines used in both works) of 0.05
dex for HR 203 and about 0.10 dex for Ti in both stars.
Apart from these deviations the agreement is excellent.  The differences
in Ti go in opposite directions for the two stars, rising doubt 
about any systematic reason. Due to the stellar parameters (low metallicity
and high temperature), any differences in the two analyses probably 
give very small abundance errors (i.e. easy continuum level definition and
spectral line fitting/measuring due to weakness of lines). 

\begin{table}
\caption{\label{tab:bdp_diff} Stars in 
common with Edvardsson et al. (1993)
(E93).}
\begin{tabular}{ l l l l l l }
\hline\noalign{\smallskip}
HR203 &  &   & HR1010 &   &  \\ 
Element & [X/H] & E93 & Element & [X/H] & E93  \\
\noalign{\smallskip}
\hline\noalign{\smallskip}
Na & -0.36 & -0.30 & Na & -0.26 & -0.22   \\	
Si &-0.21  & -0.25 & Si & -0.20 & -0.19   \\ 
Ca & -0.22 & -0.25 & Ca & -0.20 & -0.20   \\ 
Ti & -0.19 & -0.30 & Ti & -0.21 & -0.15   \\ 
Ni & -0.29 & -0.29 & Ni & -0.27 & -0.25   \\ 
Fe & -0.23 & -0.28 & Fe & -0.22 & -0.23   \\
Fe {\sc ii} & -0.26 & -0.31 & Fe {\sc ii} & -0.25 & -0.33        \\ 
      &  &  & Nd {\sc ii}  & -0.21 & -0.27\\
\noalign{\smallskip}
\hline
\end{tabular}
\end{table}

Two of the K dwarf stars, HD 32147 and HD 61606A, were also studied by
FG98. The difference in abundances are shown in 
Table \ref{tab:fg_diff}. There are large differences in the derived
abundances for several elements. For HD 32147 the reason is that the 
stellar parameters derived from the photometry were kept throughout the 
analysis for these stars in FG98, whereas here we have imposed 
excitation equilibrium in order to determine the effective temperature.
This has resulted, for HD 32147, in a change of +200 K of the effective
temperature. In FG98 it was not possible, for the
K dwarf stars, to derive effective temperatures from the excitation 
equilibrium due to a lack of useful Fe {\sc i} lines at low
excitation potential in the observed spectra. 

Spectra for five stars of the FG98 sample have been reanalyzed by us 
with the methods described above (lower part of Table 
\ref{tab:starspectable}). The parameters from FG98 were
initially used. Surface gravity was changed for some of the stars. 
The overlapping spectral regions
(only overlapping regions were used, allowing us to use the
same set of lines) provided less number of lines than the original
set  used for our twelve program stars. The regions are presented in Table 
\ref{tab:observedregions}. 
We could not justify any change in effective temperature
as for the observational sample in this work since the number of available 
low excited Fe {\sc i} lines  were too low for that decision. It seems however, 
when studying the abundance patterns and the (lack of) overionization, that
the temperatures of the stellar models are rather well chosen. 
 The resulting abundances are
presented in  the lower part of Table \ref{tab:ab_all} and the stars are 
represented in the abundance
diagrams  by circles. The differences in abundances between 
FG98 and this work for the five stars are similar to those of HD 32147 and
HD 61606A, see Sect. \ref{sect:abund} and Table \ref{tab:fg_diff}.

\begin{table*}\caption{\label{tab:fg_diff} 
 Two K dwarfs in common with Feltzing \& Gustafsson (1998).  Stellar model 
parameters used in the abundance analysis are showed, as well as the parameters
FG98 used.
 [X/H] give the abundance derived using the stellar models showed
(see also Table \ref{tab:starspectable}). The columns labeled FG98 present 
the data originally published in (appendix) Table 1 in FG98, while the 
F95Recalc column gives the values Feltzing (1995) found when increasing 
$T_{\rm eff}$ in HD 32147 by +200 K as compared to $T_{\rm eff}$ given in FG98.}
\begin{tabular}{ lrrr  lrr }
\noalign{\smallskip}
\hline\noalign{\smallskip}
HD 32147 &  &  &   &  HD 61606A &  \\ 
\multicolumn{4}{l}{$T_{\rm eff}$/log g/[Me/H]} & \multicolumn{3}{l}{$T_{\rm eff}$/log g/[Me/H]}\\
\multicolumn{4}{l}{4825 K/4.57/0.29} & \multicolumn{3}{l}{4833/4.55/0.06}\\
\multicolumn{4}{l}{4625 K/4.57/0.17 (FG98)} & \multicolumn{3}{l}{4833/4.55/0.11 (FG98)}\\
   & [X/H] & FG98 & F95Recalc  &   & [X/H] & FG98 \\
\noalign{\smallskip}
\hline\noalign{\smallskip}
Na   	&  0.46 &      &    &Na   	& 0.04 	& -0.25 \\	
Si 	& 0.44 	& 0.48 &0.31&Si 	& 0.11 	&  0.03 \\ 
Ca 	&  0.31 & 0.01 &0.18&Ca 	&  0.12 & -0.17 \\ 
Sc {\sc ii} 	& 0.44  & 0.49 &0.45&Sc {\sc ii}	& 0.16 	&  0.02 \\
Ti 	&  0.40 & 0.11 &0.34&Ti 	& 0.01  & -0.09 \\ 
V 	&  0.44 & -0.18&0.06&V 		& -0.04 & -0.46 \\ 
Cr 	&  0.32 & 0.10 &0.20&Cr 	& 0.06  & -0.10 \\ 
Cr {\sc ii} 	& 0.34  & 0.78 &$\sim$0.57&Cr {\sc ii} &  0.11 &  0.24 \\ 
Fe 	&  0.34 & 0.22 &0.19&Fe 	&  0.07 &  -0.08\\
Fe {\sc ii} 	& 0.45  & 0.61 &$\sim$0.30&Fe {\sc ii} & 0.21  &  0.09 \\ 
Co 	&  0.50 & 0.39 &0.40&Co 	& 0.01  &  -0.11\\ 
Ni 	& 0.45	& 0.57 &0.53&Ni 	& 0.05 	&  -0.03\\ 
Nd {\sc ii} 	&  0.49	&      &    &Nd {\sc ii} 	& 0.26 	&  -0.18 \\
\noalign{\smallskip}
\hline
\end{tabular}
\end{table*}

\section{Stellar abundances}
\label{sect:abund}

\begin{figure*}                                           
  \resizebox{\hsize}{!}{\includegraphics{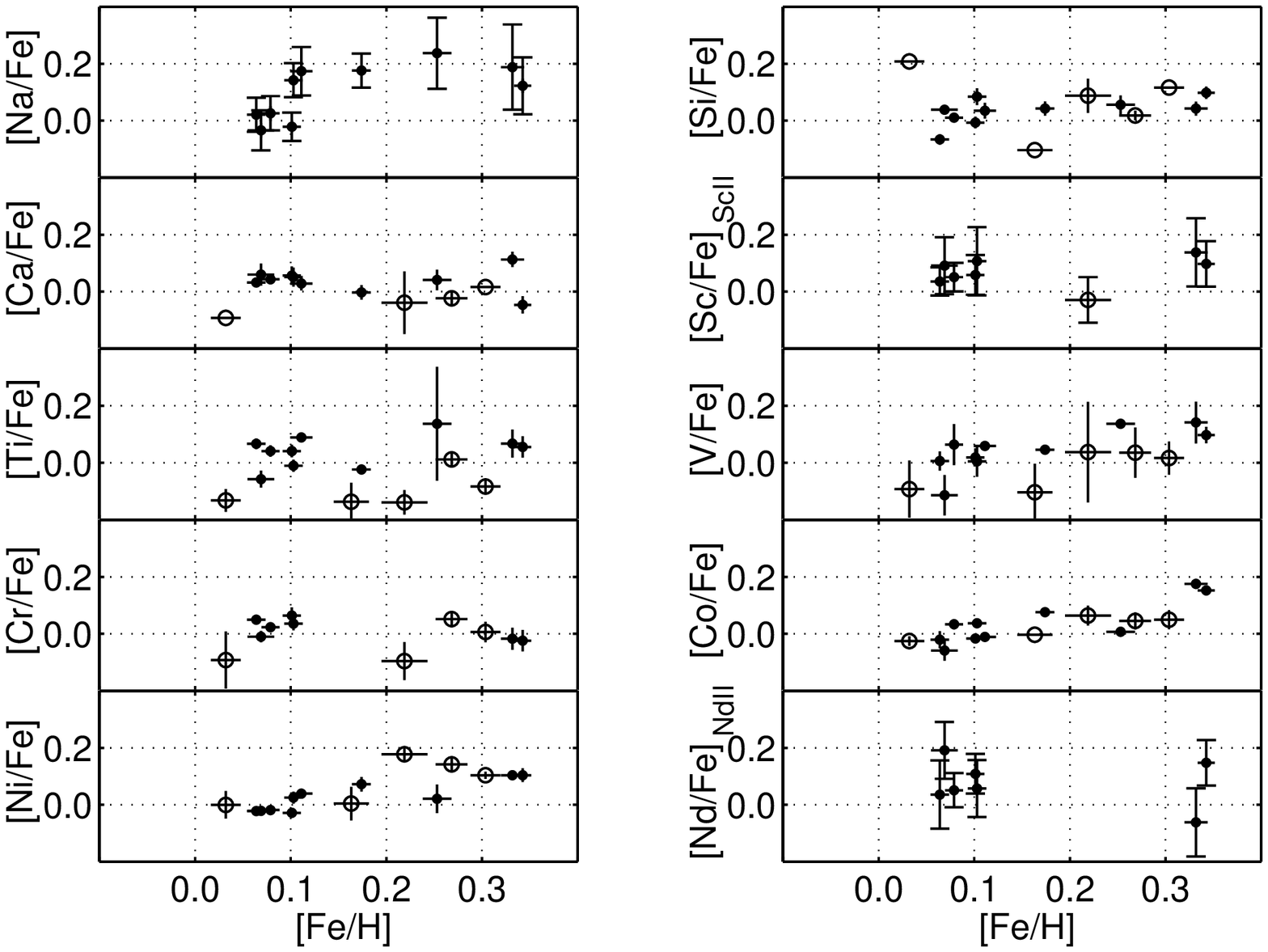}}
  \hfill
  \parbox[b]{180mm}{
    \caption{Abundance patterns. Dots : abundances for the ten  metal rich 
    stars in the sample of this work. Circles : abundances
    for the five stars of the FG98 re-analysis sample analysed with the methods
    described in this work. Error bars : formal error (the error in the mean = line-to-line 
    scatter divided by square root of (number of lines - 1) ). For the elements with only one or 
    two lines available the formal error has been replaced with mean uncertainty in  the spectral 
    fitting of the lines (error bars with 'T' at end).}
    \label{fig:all_elements}}
  \end{figure*}

We present our final stellar abundances in Table \ref{tab.abundall}
 and   (excluding the hot, metal poor dwarfs) Fig. \ref{fig:all_elements}. 

The most notable result for our cool metal-rich dwarf stars is that the
two most metal-rich stars (HD 32147 and HD 21197) show almost identical
abundance patterns for most of the elements in this study. In particular
this is true for V, Co, Sc {\sc ii}, Ti, Si, Cr, and Ni. This is an important
finding considering that FG98 found considerable
scatter for several of these elements at this metallicity ([Fe/H] = 0.30 dex). 
The reasons for this are discussed for each element separately below.

The five stars analysed with spectra from the 2.7m McDonald telescope run
made by Feltzing in 1994 are in this section referred to as the 
FG98  re-analysis sample. 
The abundance patterns of these essentially follow the ones from the stars
observed in this work but with larger scatter in many cases. For some elements
also a systematic behaviour can be observed. The conclusions of the 
results are primarily based on the ESO CAT sample.

 \begin{figure*}                                           
  \resizebox{\hsize}{!}{\includegraphics{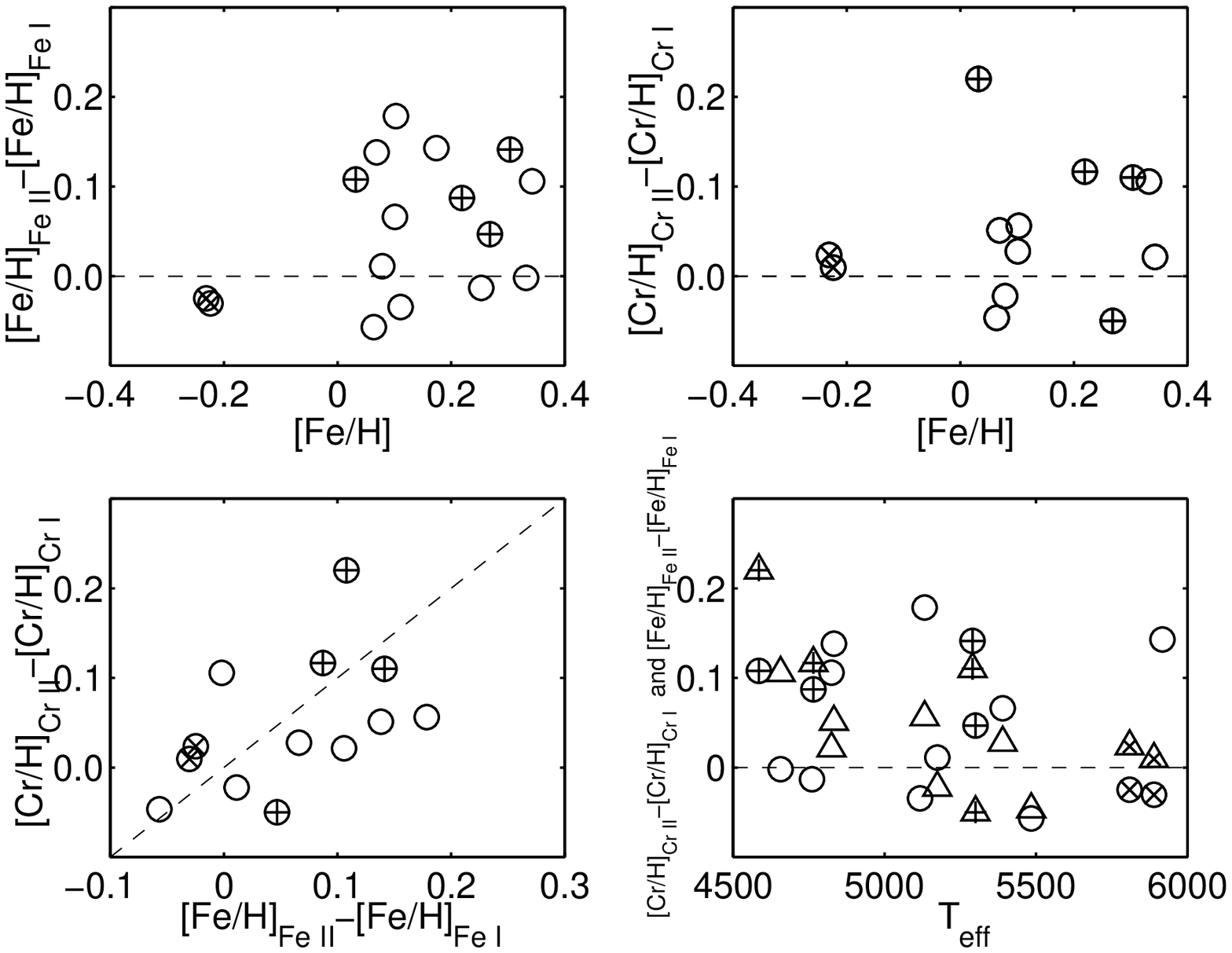}}
  \hfill
  \parbox[b]{180mm}{
    \caption{Different ionization patterns. Symbols with plusses represent
    abundances for stars from the FG98  re-analysis sample, analysed with the methods
    described in this work. Symbols with $\times$s represent abundances for the
    two hot, metal poor dwarfs. In the last plot Cr is represented
    by triangles and Fe by circles.}
    \label{fig:ion_eq}}
  \end{figure*}

\paragraph{Sodium}

[Na/Fe] increases with [Fe/H], which is also seen 
in EAGLNT93 and FG98.  The 
astrophysical oscillator strength (log $gf$) values 
used here are essentially the same as in FG98.  

\paragraph{Silicon}

There is a weak overabundance for Si but no increasing trend appears.
 This result follows that indicated by the metal rich  end in
 EAGLNT93. It does not confirm the overabundance indicated for Si in
 cool dwarf stars in FG98. 1-3 lines were
 used in FG98, 13 in this work. The
 line-to-line scatter in our data is significantly reduced compared
 with FG98. The error analysis in FG98 
 (see their Table 3 and also several tables in
 Feltzing, 1995) indicates that for cool K dwarf stars like HD
 32147 an increase in effective temperature of 200 K is followed by a
 decrease in derived Si abundance of around 0.15 dex. The $T_{\rm eff}$
 adjustment in Sect. \ref{sect:par_det_desc} lead to that the K dwarfs
 follow the same trend for Si as do the G dwarfs.

These new results show that the large scatter seen in [Si/Fe] in 
FG98 was partly due to a 
small numbers of (not optimally selected) lines as well as too low effective 
temperatures for K dwarf stars. We thus 
conclude that Si varies in lockstep with Fe also for stars more metal-rich
than 0.2 dex.

\paragraph{Calcium} 

As shown in Thor\'en (2000), the major reason for the underabundance in
 Ca for K dwarf  in FG98 was  due to
 erroneously calculated line damping parameters for the Ca lines.

Any remaining differences in [Ca/Fe] between G and K dwarf stars
vanishes when the effective temperatures are adjusted  to satisfy the
excitation balance, Sect. \ref{sect:par_det_desc}. The  resulting Ca trend
and scatter follows what is indicated in EAGLNT93
and in FG98 for the F and G dwarf stars.

\paragraph{Scandium } 

One Sc {\sc ii} line was used, four in FG98. They found that Sc {\sc ii} follows
Fe, except for the K dwarf stars, where a large underabundance can 
be seen. This is not the case here, where Sc {\sc ii} follows Fe and the warmer
dwarf stars. With only one line no strong conclusions can be drawn, though.

\paragraph{Titanium}

We find [Ti/Fe] to vary in lock step with Fe also for K dwarf
stars. This is in contrast to FG98 where the
K dwarf stars were low in Ti as compared to the G dwarf stars.  This
is understandable. Abundances derived from Ti {\sc i}
lines are sensitive to variations in effective temperature, see Table 3 in
FG98 and tables in Feltzing (1995)\nocite{feltzingthesis}.
 In fact they found that the
effects are larger on K dwarf stars than on G dwarf stars.

The difference of 200 K is enough to bring the Ti {\sc i} up 0.24
dex, and  thus move the K dwarf stars onto solar [Ti/Fe] and with a
small scatter  for our main sample. 
We note however, that the FG98 re-analysis sample shows lower [Ti/Fe].

\paragraph{Vanadium}

V is overabundant at high [Fe/H]. This is opposite to the case seen in
FG98 where the K dwarf stars show
underabundances.  Four lines are used here, two in FG98. Since we here increase the 
lowest effective temperature by 200 K, Sect. \ref{sect:par_det_desc}, we should  expect the
underabundance seen for K dwarf stars in FG98
to disappear since their error analysis (FG98, 
Table 3) indicate that an increase of +200 K will result in and V
abundance increased by 0.23 dex.  Furthermore, care has also to be
taken with broadening of the strong lines. As shown in  \thorenletter \ 
erroneous line broadening may be fatal for strong lines. The effect is
worst for the coolest, most metal rich stars.  This has been checked
for V but no effect could be seen  when changing the   damping
scheme of Anstee \& Barklem (1997), Barklem (1998) used in SPECTRUM to
the values used in FG98.  Two lines are used
in FG98, one (6452.31 {\AA}) is in common
with  this work. This line has in FG98 an
astrophysical log $gf$ value 0.5 dex higher than the one used
here. This is  probably the reason for the difference in behaviour for
V. When  the $\log gf$ value of FG98 is
inserted here and the synthetic  line fitted, it gives an
underabundance of about 0.2 dex.

Taking the five stars of the FG98  re-analysis sample into account the overabundance
is somewhat reduced. They do, however, also show an increasing trend
 with increasing [Fe/H]. 

\paragraph{Chromium}

 No sign of Cr under- or overionization is apparent Fig. \ref{fig:ion_eq}. 
 The data in Fig. \ref{fig:all_elements} follows the trend seen in  
FG98, except for the K dwarf stars which are underabundant in that
work, this is consistent with the errors analysis presented in
FG98 (Table 3) where they show that an
increase in effective temperature would result in abundances from Cr
{\sc i} lines would increase by 0.1 dex while  those from Cr {\sc ii}
would decrease by 0.15 dex, thus reducing the apparent over-ionization
drastically.

From our work it is apparent that by imposing excitation equilibrium,
see Sect. \ref{sect:par_det_desc}, we naturally arrive close to ionization
equilibrium as well.

\paragraph{Iron}

A weak Fe overionization (mean $<$ 0.1 dex), with some scatter is seen in
Fig. \ref{fig:ion_eq}.  The effect seen for K dwarf stars in  
FG98 is mostly removed though, and no temperature  trend for
Fe {\sc ii} can be observed, as seen in  
Fig. \ref{fig:ion_eq}.

\paragraph{Cobalt}

An increasing trend with [Fe/H] and small scatter is apparent for Co. This can  be
compared to FG98 where an increasing trend
is seen for the K dwarfs but not for the other stars. The overabundance 
here is $\sim$ 0.2 dex for the most metal rich dwarf stars. Eight lines were 
used in FG98, four lines here.  Both their K dwarf stars and the
dwarf stars here appear to follow the upper  envelope of the slowly
increasing [Co/Fe] trend for the whole sample of FG98.
The five stars from the FG98  re-analysis sample moderate the trend somewhat.

\paragraph{Nickel}

FG98 found that Ni varied in lock-step with
Fe except the K dwarf stars, which were more abundant in Ni than the G
dwarf stars. We show that also the K dwarf stars indeed vary in
lock-step with iron also for high metallicities. Taking the five stars
from the FG98 sample into account a weak increasing trend is revealed. 
 Part of the decreased [Ni/Fe] here as compared to FG98
for the K dwarf stars can be attributed to the change in effective
temperatures, Sect. \ref{sect:par_det_desc}, but not all. Some of the
stars in FG98 showed a considerable scatter
in [Ni/Fe] perhaps indicating problems with the analysis of some of
the lines used. All of their K dwarf stars, except HD 61606A, show such
large scatter, and thus we conclude that part of the high [Ni/Fe]
found for the K dwarf stars in their work was also due to problems
with the analysis of some of the Ni lines.

We note, however, that the two most metal-rich stars in our sample
are slightly more abundant in Ni then the rest of the sample. The
new re-analyzed spectra from the FG98 sample show a weak increasing trend with
increasing [Fe/H]. They additionally show a larger scatter.

\paragraph{Neodynium}
We only use one line for Nd, not the same as the single line used in 
FG98, and the trend seen in their work is not reproduced here. Rather, Nd seems 
to follow iron, with a considerable scatter. However, any firm conclusion
from this material is impossible.

\begin{table*}
\caption{\label{tab:ab_all} Stellar abundances.  The second section
shows the two hot, metal poor dwarfs. The last five stars represent
the FG98  re-analysis sample (objects observed 1994 at the 2.7m McDonald telescope, Sect. \ref{sect:instrument}).}
\begin{tabular}{lrrrrrrrrrrrrrrrrrrr}
\noalign{\smallskip}
\hline\noalign{\smallskip}
         &     Na & Si & Ca & Sc {\sc ii} & Ti  & V & Cr {\sc i} & Cr {\sc ii} & Fe {\sc i} & Fe {\sc ii} & Co & Ni  & Nd {\sc ii} \\ 
\noalign{\smallskip}
\hline\noalign{\smallskip}
HD 12235 &  0.35 & 0.22 & 0.17 &   & 0.15 & 0.22 &   &   & 0.17 & 0.32 
& 0.25 & 0.25 &   \\ 
HD 21197 &  0.52 & 0.37 & 0.44 & 0.47 & 0.40 & 0.47 & 0.31 & 0.42 & 0.33 & 0.33 
& 0.51 & 0.44 & 0.27 \\ 
HD 23261 &  0.24 & 0.19 & 0.15 & 0.21 & 0.09 & 0.11 & 0.14 & 0.20 & 0.10 & 0.28 
& 0.14 & 0.13 & 0.16 \\ 
HD 30501 &  0.11 & 0.09 & 0.12 & 0.13 & 0.12 & 0.14 & 0.10 & 0.08 & 0.08 & 0.09 
& 0.11 & 0.06 & 0.13 \\ 
HD 31392 &  0.08 & 0.09 & 0.16 & 0.16 & 0.14 & 0.12 & 0.17 & 0.19 & 0.10 & 0.17 
& 0.08 & 0.07 & 0.21 \\ 
HD 32147 &  0.46 & 0.44 & 0.30 & 0.44 & 0.40 & 0.44 & 0.32 & 0.34 & 0.34 & 0.45 
& 0.50 & 0.45 & 0.49 \\ 
HD 42182 &  0.29 & 0.15 & 0.14 &   & 0.20 & 0.17 &   &   & 0.11 & 0.08 
& 0.10 & 0.15 &   \\ 
HD 61606A &  0.03 & 0.11 & 0.13 & 0.16 & 0.01 &-0.05 & 0.06 & 0.11 & 0.07 & 0.21 
& 0.01 & 0.05 & 0.26 \\ 
HD 69830 &  0.08 & 0.00 & 0.10 & 0.10 & 0.13 & 0.07 & 0.11 & 0.07 & 0.06 & 0.01 
& 0.04 & 0.04 & 0.10 \\ 
HD 213042 &  0.49 & 0.31 & 0.29 &   & 0.39 & 0.39 &   &   & 0.25 & 0.24
 & 0.26 & 0.27 &   \\ 
 \noalign{\smallskip}
\hline\noalign{\smallskip} 
\noalign{\smallskip}
HR 203 & -0.36 &-0.21 &-0.22 &-0.18 &-0.19 &-0.28 &-0.20 &-0.18 &-0.23 &-0.26 &-
0.19 &-0.29 &-0.18 \\ 
HR 1010 & -0.26 &-0.20 &-0.20 &-0.26 &-0.21 &-0.26 &-0.25 &-0.24 &-0.22 &-0.25 &
-0.23 &-0.27 &-0.21 \\ 
 \noalign{\smallskip}
\hline\noalign{\smallskip} 
\noalign{\smallskip}
HD 77338  &   &  0.42 & 0.32 &   & 0.22 & 0.32 & 0.31 & 0.42 & 0.30 & 0.44 & 0.35 & 0.41 &   \\ 
HD 87007  &   &  0.29 & 0.25 &   & 0.28 & 0.30 & 0.32 & 0.27 & 0.27 & 0.32 & 0.31 & 0.41 &   \\ 
HD 103932 &   &  0.06 &   &   & 0.03 & 0.06 &   &   & 0.16 &   & 0.16 & 0.17 &   \\ 
HD 131977 &   &  0.24 &-0.06 &   &-0.10 &-0.06 &-0.06 & 0.16 & 0.03 & 0.14 & 0.01 & 0.03 &   \\ 
HD 136834 &   &  0.31 & 0.18 & 0.19 & 0.08 & 0.26 & 0.12 & 0.24 & 0.22 & 0.31 & 0.28 & 0.40 &   \\ 
 \noalign{\smallskip}
\hline
\end{tabular}
\label{tab.abundall}
\end{table*}

\subsection{Summary and speculations}

We show that [Si/Fe], [Ca/Fe], [Ti/Fe], and [Cr/Fe] all
varies as [Fe/H] between 0.00 and 0.35 dex. So does [Ni/Fe] but 
with a, very, tentative, upturn for the most metal-rich stars.
[Co/Fe] and [V/Fe] increase slowly with [Fe/H].
 [Na/Fe] shows a definite increasing trend with [Fe/H].
Taking the results from new observed spectra of the FG98 sample (which has
sparser amount of available lines)
into account, the [Ni/Fe] increment looks stronger and
the [Co/Fe] less convincing. 

For Nd the data is uncertain.

We find that the $\alpha$-elements such as Si and Ca follow Fe for
high metallicities, in contrast to oxygen which is indicated to keep
decreasing  with increasing [Fe/H] (FG98).  
This is, perhaps, a first indication of second sources of $\alpha$-elements
that start to contribute to the galactic chemical evolution on a very
long time-scale. This is, of course, a highly speculative suggestion
and must await further studies of in particular oxygen to be confirmed
or rejected.

%
%
\section{Discussion}
\label{sect:disc}

For the analysed elements, except V and Co and possibly Ni,
the cool dwarf stars show
the same behaviour as the warmer ones in the [A/A] diagrams. 
A weak dependence with [Fe/H] is sometimes visible but the variations 
are within the uncertainties.
Co and V show a different trend compared to FG98.  
V here increases with metallicity, while FG98 shows
a decrease. The reason may be a different (possibly erroneous) $\log gf$ value for
one of the lines used in FG98. 
Co is here also weakly increasing while no trend (although a weak, almost 
insignificant trend is seen for the K dwarf stars) is apparent in FG98. The 
cool dwarf stars in this work are situated at the upper envelope of the pattern
indicated there. Fig. 21 in FG98 shows [Co/Fe] compared 
with data from other works. A possible increase for increasing 
[Fe/H] is indeed possible to see when taking the more
metal poor stars into account. 

No evidence can be found that cool dwarf stars show non-solar
abundance ratios, other than Na (and with weak trends for Co, V 
and possibly Ni). Na is known to increase compared to Fe 
for increasing metallicity. This behaviour is reproduced
here, with a strong trend with increasing metallicity.

Cool metal rich dwarf stars seem to be usable in 
investigations of Galactic chemical evolution, if 
synthetic spectral analysis is used, rather than
equivalent widths analysis. The radiation
fields shown by the modern MARCS code indicate that only weak NLTE effects
should be expected also for other elements than Ca.

\section{Conclusions}
\label{sect:conclusion}

The investigation by FG98 left several unexpected results and unanswered
questions in particular about elemental abundances in K dwarf stars.
Applying new methods, tailored to suite the K dwarf stars,
 in our abundance analysis we have been able to answer a number of these questions. 
In particular we find that:

- by imposing excitation balance for Fe {\sc i} lines we naturally
arrive close to LTE ionization balance for Fe and Cr

- stellar abundance results for cool K dwarf stars are more or less 
the same as for their
warmer counter-parts

- we show that the stellar atmosphere models of the coolest stars need 
temperature corrections as compared to the calibrations presented by \ols. This may
indicate a systematic error in the calibration for the stars with colours $b-y>0.51$

- according to the photometric calibration for the metallicity \cite{olsen:84}, 
HD 42182 is a Super-Metal-Rich (SMR, [Me/H]$>$0.20, Taylor, 1996\nocite{taylor96}) candidate star. This is not confirmed by our data.
Also, HD 21197, which has just above solar metallicity 
from the photometry is showed to be a SMR star, as well as HD 32147

- the two most metal-rich stars in our sample show 
extremely similar abundances, which makes it possible to draw strong
conclusions on the stellar abundance trends

- [Si/Fe], [Ca/Fe], [Ti/Fe] and [Cr/Fe]
all vary as [Fe/H] also for stars more metal-rich than 0.2 dex

- [Na/Fe] (and possibly [V/Fe], [Co/Fe] and [Ni/Fe]) increases with increasing [Fe/H].

\begin{acknowledgement}
The authors thank Docent Bengt Edvardsson for valuable discussions and 
comments on the manuscript. The referee is thanked for suggestions that
improved the quality of the article. 
PT was supported by the Swedish Natural Science Research Council, NFR.
\end{acknowledgement}

\appendix

\section{Atomic line data for the LTE analysis}
\label{sect:ltedata}

The following columns appear in Table \ref{tab:linetable}:
$\chi$ is the excitation energy of the lower level of the line transition.
log $gf$ is the logarithm of the lines astrophysical oscillator strength
 times the statistical weight of the lower level of the line transition. 
 For Ca  (except the 5867 \AA \ and 6798 \AA \ lines) the log $gf$ values were taken from \smith \ and
Smith \& O'Neill (1975).
$\delta\Gamma{_6}$ represents the value used for pressure line wing damping. 
Pressure line wing damping was calculated according to 
\ansteebarklem \  for the lines marked with asterisks.
For the lines with a $\delta \Gamma{_6}$ value, this was multiplied with the
classical Uns\"old damping value. The $\delta\Gamma{_6}$ values used 
were 2.50 for all elements except Ca \cite{smith_sunproc:81,oneillsmith:80}, Na ($\delta\Gamma_6$=2.00),
 Si ($\delta\Gamma_6$=1.30) and Fe ($\delta\Gamma_6$=1.40), 
see EAGLNT93 and references therein. 
$\Gamma_{\rm rad}$ represents the radiative damping constant. The strong lines printed
in {\bf boldface} have not been used in abundance analysis, only for estimating
the surface gravity, see Sect. \ref{sect:par_det_desc}.

\begin{table}\caption{\label{tab:linetable} Atomic line data.}
\begin{tabular}{l l r l l}
\hline\noalign{\smallskip}
  Wavelength & $\chi$  & log $gf$   & $\delta\Gamma_6$  & $\Gamma_{\rm rad}$   \\
 (\AA)       &    (eV) &            &                     & (s$^{-1}$) \\
\noalign{\smallskip}
\hline\noalign{\smallskip}
  Na        &          &          &         &              \\
  6154.226  &   2.102  &  -1.650  & 2.00 & 7.079E+07 \\           
  6160.747  &   2.104  &  -1.320  & 2.00 & 7.079E+07 \\          
  Si        &          &          &         &            \\
  5873.764  &   4.930  &  -3.010  &   1.30 & 1.000E+05 \\
  6106.608  &   5.614  &  -2.230  &   1.30 & 1.000E+05 \\
  6145.016  &   5.616  &  -1.470  &   1.30 & 1.000E+05    \\
  6131.768  &   5.082  &  -3.650  &   1.30 & 1.000E+05    \\  
  6112.928  &   5.616  &  -2.250  &   1.30 & 1.000E+05  \\
  6125.021  &   5.614  &  -1.630  &   1.30 & 1.000E+05 \\
  6433.457  &   5.964  &  -1.640  &   1.30 & 1.000E+05  \\
  6800.596  &   5.964  &  -1.740  &   1.30 & 1.000E+05 \\
  8492.077  &  5.863   &  -1.930  &   1.30 & 1.000E+05 \\
  8501.544  &   5.871  &  -1.280  &   1.30 & 1.000E+05 \\  
  8502.219  &   5.871  &  -0.910  &   1.30 & 1.000E+05 \\
  8536.164  &   6.181  &  -0.610  &   1.30 & 1.000E+05 \\ 
  8556.777  &   5.871  &  -0.430  &   1.30 & 1.000E+05 \\              
  Ca        &          &          &        &            \\
  5260.387  &   2.521  &  -1.719  &   0.91 & 7.980E+07 \\
  5867.562  &   2.933  &  -1.641  &   1.01 & 2.624E+08   \\
  6161.297  &   2.523  &  -1.266  &   1.64 & 1.879E+07 \\         
\bf 6162.173 & \bf   1.899   & \bf   -0.09  & \bf    2.48 & \bf  7.244E+07 \\
  6166.439  &   2.521  &  -1.142  &   1.64 & 1.858E+07  \\             
  6169.042  &   2.523  &  -0.797  &   1.64 & 1.977E+07 \\             
  6169.563  &   2.526  &  -0.478  &   1.64 & 1.875E+07 \\  
 \bf  6439.075  &  2.526 & \bf     0.390  & \bf  0.65 & \bf 4.457E+07 \\
  6455.598  &   2.523  &  -1.290  &   0.71 & 4.645E+07 \\
  6471.662  &   2.526  &  -0.686  &   0.63 & 4.416E+07 \\
  6798.467  &   2.709  &  -2.520  &    \ \ * & 1.941E+07  \\
  Sc {\sc ii}      &          &          &         &            \\
  5239.813  &   1.455  &  -0.870  &   2.50 & 1.330E+08 \\
  Ti        &          &          &         &          \\
  5219.702  &   0.021  &  -2.392  &    \ \ * & 6.668E+06 \\
  5222.674  &   2.085  &  -0.621  &    \ \ * & 6.026E+07 \\
  5224.305  &   2.134  &  -0.210  &    \ \ * & 6.012E+07 \\
  5247.289  &   2.103  &  -0.927  &    \ \ * & 6.012E+07 \\
  5248.383  &   1.879  &  -1.418  &    \ \ * & 2.312E+08 \\
  5265.964  &   1.887  &  -0.717  &    \ \ * & 2.312E+08 \\
  5866.451  &   1.067  &  -0.940  &    \ \ * & 4.395E+07 \\
  6091.171  &   2.267  &  -0.473  &   \ \ *  & 8.492E+07 \\
  6092.792  &   1.887  &  -1.379  &   \ \ *  & 1.271E+08 \\
  6098.658  &   3.062  &  -0.210  &   \ \ *  & 5.433E+07 \\
  6121.001  &  1.879   &  -1.522  &   \ \ *  & 1.230E+08 \\
  6126.216  &  1.067   &  -1.475  &   \ \ *  & 9.931E+06 \\
  8518.352  &  1.879   &  -1.089  &    \ \ *  & 7.211E+07 \\
  V         &          &          &         &                 \\
  5240.862  &   2.374  &   0.230  &    \ \ * & 6.871E+07 \\
  6090.214  &   1.081  &  -0.062  &    \ \ * & 3.981E+07 \\
  6111.645  &   1.043  &  -0.915  &    \ \ * & 3.899E+07 \\
  6452.341  &   1.195  &  -1.306  &    \ \ * & 3.981E+07 \\
\noalign{\smallskip} 
\hline
           &           &          &           &           \\
           &           &          &           &           \\
           &           &          &           &           \\
           &           &          &           &           \\
           &           &          &           &           \\
           &           &          &           &           \\
\end{tabular}
\end{table}
\normalsize

\begin{table}
\begin{tabular}{l l r l l}
\noalign{\smallskip}
\noalign{\smallskip}
\noalign{\smallskip}
\noalign{\smallskip}
\noalign{\smallskip}
\noalign{\smallskip}
\noalign{\smallskip}
\hline\noalign{\smallskip}
 Wavelength & $\chi$  & log $gf$   & $\delta\Gamma{_6}$  & $\Gamma_{\rm rad}$   \\
 (\AA)      &    (eV) &            &                     &     (s$^{-1}$) \\
\noalign{\smallskip}
\hline\noalign{\smallskip}
  Cr        &          &          &         &           \\
  5220.912  &   3.385  &  -1.034  &   2.50 & 8.110E+07 \\
  5224.069  &   3.410  &  -1.129  &   2.50 & 8.054E+07 \\
  5225.814  &   2.708  &  -1.519  &    \ \ * & 3.622E+08 \\
  5228.096  &   3.369  &  -0.808  &   2.50 & 1.361E+07 \\
  5238.964  &   2.709  &  -1.505  &    \ \ * & 2.541E+08 \\
  5240.464  &   3.668  &  -0.804  &   2.50 & 4.688E+07 \\
  5243.364  &   3.395  &  -0.667  &   2.50 & 8.091E+07 \\
  5247.566  &   0.961  &  -1.740  &    \ \ * & 5.284E+07 \\
  5265.157  &   3.428  &  -0.519  &   2.50 & 8.091E+07 \\
  Cr {\sc ii}     &          &          &         &           \\
  5237.329  &   4.073  &  -1.377  &   2.50 & 2.553E+08 \\
  5246.768  &    3.714 &   -2.566 &   2.50 & 2.312E+08 \\
  5249.437  &   3.758  &  -2.789  &   2.50 & 2.350E+08 \\
  Fe        &          &          &         &           \\
\bf 5227.151 & \bf 2.424 & \bf -1.352  &  \ \ * & \bf 1.449E+08 \\
\rm  5837.701  &   4.294  &  -2.340  &    \ \ * & 4.775E+07 \\
  6078.491  &   4.795  &  -0.374  &   \ \ *  & 1.828E+08 \\
  6079.009  &   4.652  &  -1.020  &   \ \ *  & 1.932E+08 \\
  6082.711  &   2.223  &  -3.573  &   \ \ *  & 7.691E+06 \\
  6089.580  &   4.580  &  -1.312  &   1.40 & 9.268E+07 \\
  6093.644  &   4.607  &  -1.400  &  \ \  *  & 1.936E+08 \\
  6093.644  &   4.607  &  -1.400  &   1.40 & 1.936E+08 \\
  6094.374  &   4.652  &  -1.640  &  \ \  *  & 1.932E+08 \\
  6096.665  &   3.984  &  -1.770  &  \ \  *  & 4.529E+07 \\
  6098.245  &   4.558  &  -1.840  &  \ \  *  & 2.618E+08 \\
  6105.131  &   4.548  &  -2.050  &  \ \  *  & 1.941E+08 \\
  6120.249  &   0.915  &  -5.950  &  \ \  *  & 2.710E+04 \\
  6145.420  &   3.368  &  -3.600  &  \ \  *  & 1.175E+08  \\
  6151.618  &   2.176  &  -3.379  &  \ \  *  & 1.549E+08 \\ 
  6157.728  &   4.076  &  -1.320  &  \ \  *  & 5.023E+07 \\           
  6159.378  &   4.607  &  -1.970  &  \ \  *  & 1.919E+08 \\
  6165.360  &   4.143  &  -1.554  &  \ \  *  & 8.770E+07 \\
  6173.336  &   2.223  &  -2.920  &  \ \  *  & 1.671E+08 \\
  6180.204  &   2.727  &  -2.686  &   1.40 & 1.469E+08 \\
  6187.39   &   2.832  &  -4.339  &  \ \  *  & 1.449E+08 \\
  6430.846  &   2.176  &  -2.006  &   1.40 & 1.648E+08 \\
  6436.407  &   4.186  &  -2.410  &    1.40 & 3.041E+07 \\
  6469.193  &   4.835  &  -0.770  &    1.40 & 2.275E+08 \\
  6481.870  &   2.279  &  -2.984  &    1.40 & 1.549E+08 \\
  6752.707  &   4.638  &  -1.204  &    1.40 & 2.301E+08\\
  6756.563  &   4.294  &  -2.750  &   \ \  *  & 7.345E+07 \\
  6786.860  &   4.191  &  -1.850  &    1.40 & 1.986E+08 \\
  6794.619  &   4.955  &  -2.110  &    1.40 & 4.355E+08  \\
  6804.001  &   4.652  &  -1.546  &    1.40 & 1.758E+08 \\
  6804.271  &   4.584  &  -1.813  &    1.40 & 5.236E+07 \\
  6806.845  &   2.727  &  -3.110  &    1.40 & 1.021E+08\\
  6810.263  &   4.607  &  -0.986  &    1.40 & 2.301E+08 \\
  6819.586  &   4.103  &  -2.677  &   \ \  * & 2.128E+08 \\
  6820.372  &   4.638  &  -1.120  &    1.40 & 2.218E+08 \\
  8471.739  &   4.956  &  -0.963  &    1.40   &  4.710E+08  \\
  8481.982  &   4.186  &  -1.981  &    1.40  & 6.295E+07 \\
  8514.072  &   2.198  &  -2.129  &   \ \ * & 1.959E+07  \\
  8515.108  &   3.018  &  -1.973  &   1.40  &  7.568E+07  \\
\noalign{\smallskip} 
\hline
\end{tabular}
\end{table}
\normalsize

\begin{table}
\begin{tabular}{l l r l l}
\noalign{\smallskip}
\noalign{\smallskip}
\noalign{\smallskip}
\noalign{\smallskip}
\noalign{\smallskip}
\noalign{\smallskip}
\noalign{\smallskip}
\hline\noalign{\smallskip}
 Wavelength & $\chi$  & log $gf$   & $\delta\Gamma{_6}$  & $\Gamma_{\rm rad}$   \\
 (\AA)      &    (eV) &            &                     &     (s$^{-1}$) \\
\noalign{\smallskip}
\hline\noalign{\smallskip}
   Fe {\sc ii}      &         &       &         &               \\
  6084.111  &   3.199  &  -3.800  &    2.50 & 3.428E+08 \\
  6113.322  &   3.221  &  -4.210  &    2.50  & 3.412E+08 \\
  6149.258  &   3.889  &  -2.870  &   \ \  *  & 3.388E+08 \\
  6432.680  &   2.891  &  -3.670  &    2.50 & 2.897E+08 \\
  6456.383  &   3.903  &  -2.250  &    2.50 & 3.373E+08  \\
  Co        &          &          &          &           \\ 
  6093.143  &   1.740  &  -2.440  &  \ \   * & 2.080E+07 \\
  6116.996  &   1.785  &  -2.590  &  \ \   * & 2.009E+07 \\
  6454.990  &   3.632  &  -0.350  &  \ \   * & 7.396E+07 \\
  Ni        &          &          &         &            \\
  6108.107   &  1.676  &  -2.450  &   \ \  * & 4.864E+07 \\
  6111.066   &  4.088  &  -0.870  &  \ \   * & 1.455E+08 \\
  6119.749   &  4.266  &  -1.350  &    2.50 & 2.673E+08 \\
  6128.963   &  1.676  &  -3.430  &   \ \  * & 1.211E+07 \\
  6133.963   &  4.088  &  -1.830  &   \ \  * & 1.449E+08 \\
  6175.360   &  4.089  &  -0.559  &   \ \  * & 2.328E+08 \\
  6176.807   &  4.088  &  -0.350  &   \ \  * & 1.452E+08 \\
  6177.236   &  1.826  &  -3.600  &    2.50 & 4.305E+07 \\
  6183.842   &  4.167  &  -1.938  &    2.50 & 2.529E+08 \\
  6186.709   &  4.105  &  -0.960  &  \ \   * & 2.056E+08 \\
  6482.796   &  1.935  &  -2.930  &  \ \   * & 8.147E+07 \\
  6772.313   &  3.658  &  -0.980  &  \ \   * & 1.500E+08 \\
\noalign{\smallskip} 
\hline
\end{tabular}
\end{table}
\normalsize

\end{document}